\documentclass[11pt, oneside]{article}   	
\usepackage{geometry}                		
\usepackage{graphicx}				
\usepackage{amssymb}

\title{Quantifying Absorption in the Transactional Interpretation}
\author{R. E. Kastner\footnote{Foundations of Physics Group, University of Maryland, College Park; rkastner@umd.edu},\   John G. Cramer\footnote{Department of Physics, University of Washington}}

\date{\today}							

\begin{document}
\maketitle


\begin{abstract}

The Transactional Interpretation offers a solution to the measurement problem by identifying specific physical conditions precipitating the non-unitary `measurement transition' of von Neumann. Specifically, the transition occurs as a result of absorber response (a process lacking in the standard approach to the theory).  The purpose of this Letter is to make clear that, despite recent claims to the contrary, the concepts of `absorber' and `absorber response,' as well as the process of absorption, are physically and quantitatively well-defined in the transactional picture. In addition, the Born Rule is explicitly derived for radiative processes. 

\end{abstract}
\maketitle

\section{Introduction and Background}

\indent The Transactional Interpretation (TI) \cite{Cramer1},\cite{Cramer2}, is based on the direct-action theory of electromagnetism by Wheeler and Feynman (WF)\cite{WF}. The fully relativistic version of TI \cite{Kastner2012} is based on Davies' direct-action theory of quantum electrodynamics\cite{Davies71},\cite{Davies72}. The Davies theory proposes, in analogy with the classical Wheeler-Feynman theory, that the basic field interaction is a direct connection between charges, and is a time-symmetric one (rather than future-directed). The causal or future-directed behavior of the observable field--i.e., the field that conveys a real photon from one charge to another--then derives from the response of absorbers. 

\indent This section offers a brief review of the basic transactional picture. However, it is assumed that the reader is familiar with the basics of TI, which can be found in the above references (see also \cite{Kastner2018}).The following section demonstrates that the quantum relativistic level of the interpretation (RTI) provides for quantification and precise definition of the concept of `absorber response,' thus refuting claims that `absorber' and `absorption' are not appropriately defined in TI (e.g., \cite{Marchildon}). It also presents a derivation of the Born Rule for radiative processes, one which is available only in the direct-action theory of fields.\footnote{It should be noted that RTI involves real dynamics and denies the usual `block world' ontology that is often presupposed in connection with `retrocausal' interpretations. In RTI, the formal time symmetry of the basic field propagation does not equate to time symmetry at the spacetime level. Due to space considerations, we do not discuss those ontological details here, but refer the interested reader to relevant publications such as \cite{Kastner Causal Sets}, \cite{Kastner Kauffman Epperson}, \cite{Kastner Entropy}. We note here that the RTI ontology involves actualization of possibles; that is what generates the requirement (pertaining only to quantum fields as opposed to classical fields) to multiply amplitudes to obtain a probability. Measurement outcomes are not simply given in a static block world, but are truly ontologically uncertain, and that uncertainty is quantified by the product of the amplitudes for emission and absorption, which are both needed in order to establish the invariant spacetime interval corresponding to a given measurement outcome.}

\indent First, an `absorber' in TI is simply an elementary bound system such an atom or molecule capable of being excited into a higher internal state; i.e. an object understood as an absorber in standard physics. In the transactional picture, the usual quantum state vector or `ket' $| \Psi \rangle$ is called an `offer wave' (OW), and the advanced response or dual vector $\langle a|$ of an absorber $A$ is called a `confirmation wave' (CW). The absorber also generates a time-symmetric field, but one that is exactly out of phase ($e^{i\pi}$) with the field received by it.  Under these conditions, propagation to the future of the absorber and to the past of the emitter is cancelled, and the retarded field between the emitter and absorber is reinforced (\cite{Cramer1}). The result is that a future-directed real field arises between the emitter and absorber, constituting a quantum of excitation of the electromagnetic field.  

In general, many absorbers respond to an OW. Each absorber responds to the momentum component of the OW that reaches it. For example, consider the typical situation of an excited atom surrounded by \textit{N}  ground state atoms $G_{i}, i=\{1, N\}$. The decay of the excited atom yields an offer wave $|\Psi\rangle$, which is a normalized sum over plane waves of momentum $\vec{k}_i$. Ground state atoms $G_i$, receiving the component $\langle \vec{k_i} | \Psi \rangle | \vec{k_i} \rangle $ from the emitter, generate their own time-symmetric field, whose advanced components propagate back to the emitter. Each $G_i$'s  advanced response is represented by the dual vector  $\langle  \Psi | \vec{k_i} \rangle   \langle \vec {k_i}|$. 

The product of the amplitudes of the OW and CW components clearly corresponds to the Born Rule. The relevance of their product is that this describes the final amplitude after a complete `circuit' from emitter to absorber and back again; this was shown in Cramer (1986)\cite{Cramer1}. For a particular responding atom $G_m$, the outer product of the entire OW and CW yields a weighted projection operator,  

$$| \langle \vec{k_m} | \Psi \rangle|^2  | \vec{k_m}\rangle \langle \vec{k_m} |  \eqno(1)$$
where the weight is the Born Rule. Taking into account the responses from all $G_i$, we get a sum of the weighted outer products corresponding to all CW responses:

$$ \sum_i  | \langle \vec{k_i} | \Psi \rangle|^2  | \vec{k_i}\rangle \langle \vec{k_i} |  \eqno(2)$$

 This constitutes the mixed state identified by von Neumann as resulting from the non-unitary process of measurement (cf Kastner (2012), Chapter 3).\cite{Kastner2012}. This is the manner in which TI provides a physical explanation for both the Born Rule and the measurement transition from a pure to a mixed state. I.e, unitarity is broken upon the generation of CWs as above, since this process transforms the state vector to a convex sum of weighted projection operators. 

The weighted projection operators for the outcomes, i.e., the components of the density matrix resulting from measurement, represent \textit{ incipient transactions}. At this point, an unstable situation is set up, since there is only one photon, whose conserved quantities can only be received by one (not all) of the responding absorbers. Thus, indeterministic, non-unitary collapse occurs. (The additional step from the mixed state to the `collapse' to just one outcome is understood in RTI as an analog of spontaneous symmetry breaking.) The `winning' transaction, corresponding to the outcome of the measurement, is termed an \textit{actualized transaction}, and the absorber that actually receives the quantum is called the \textit{receiving absorber}. Thus, in general, many other absorbers participate in the process by responding with CW (thus canceling the emitted components not ultimately actualized/absorbed), but do not end up receiving the actualized quantum. 

The upshot of the above is that once an atom transitions from its ground state $|G\rangle$ to an excited stationary state $|E\rangle$, it has definitely `absorbed' a photon, and linearity of the photon state propagation has been broken, since there has been a \textit{physical} transition to a sum over projection operators $| \vec{k_i}\rangle \langle \vec{k_i} | $ as above, and collapse to an outcome $k_m$ represented by a single projection operator $| \vec{k_m}\rangle \langle \vec{k_m} | $. Since we can confirm these sorts of state transitions (i.e., one can detect whether atoms are in ground or excited states, or some other arbitrary state), we can pinpoint at what stage in an interaction a non-unitary transition has taken place. But actually, the result of including absorber response as a physical process is a stronger one: TI \textit{predicts that we will find distinct outcomes corresponding to such transitions}--which we do--rather than having to take our experience of finding such distinct outcomes as in need of explanation (the latter being the case under an assumption of continuing linear evolution.) 

We make this more quantitative in the next section.

\section{Quantum relativistic treatment}

\indent  As noted above, Davies provided a quantum relativistic version of the Wheeler-Feynman theory. In the Davies theory, the basic electromagnetic field $A^\mu$ is non-quantized and the basic field propagation is represented by the time-symmetric propagator, rather than by the usual `causal' Feynman propagator of standard quantum electrodynamics (QED). (Instead, `causal' behavior is derived from the responses of absorbers, rather than needing to be postulated separately.) The time-symmetric propagator can be defined in terms of the retarded and advanced propagators (Green's functions) of the electromagnetic field, i.e.:

$$\bar{D} = \frac{1}{2} D_{ret} + \frac{1}{2} D_{adv} \eqno(3) $$

In analogy with the Wheeler-Feynman theory, Davies shows that response from other charges to the above time-symmetric field from a given charge results in the Feynman causal propagator, which includes a term corresponding to radiation, i.e., the emission of real photons. This can be seen by looking at the Feynman propagator in momentum space (ignoring metric factors):

$$ D_F(k) =  \frac{1}{k^2 + i\epsilon} = P(\frac{1}{k^2}) - i \pi \delta (k^2)  \eqno (4) $$

\noindent where $P(\frac{1}{k^2})$ denotes the principal value (i.e., the pole is excluded), and the delta function corresponds to the pole and therefore represents a real photon (i.e., a photon `on the energy shell,' such that it has zero rest mass, as opposed to a virtual photon (see Davies 1972, p. 1027).  Upon transforming to coordinate space, the first term corresponds to $\bar{D}$ and the delta function term corresponds to the `free field' Green's function $D_1$, a solution to the homogeneous equation. 

Radiated real photons correspond to Fock states having precise photon number. As in standard QED, they are transverse only, and are quantized, corresponding to the $D_1$ component. However, the time-symmetric component $\bar D$ (obtaining in the absence of absorber response) is not. Virtual photons correspond to the time-symmetric propagator  (i.e., propagation not prompting absorber response), and therefore are not quantized. (This feature provides a resolution to the consistency problems facing interacting quantum field theories, such as Haag's Theorem.\cite{KastnerHaag})

In the Davies theory, the usual quantum electromagnetic field $A(x)$ (suppressing component indices for simplicity, i.e. $A=A^\mu$ and $x=x^\mu$) is replaced by the direct current-to-current interaction as above. Thus, the field at a point $x$ (on a charged current $i$) arising from its interaction with responses of all currents $j_j$ is given by 

$$ A(x) =  \sum_j  \int D_F (x-y) j_{j}(y) d^4y \eqno(5) $$

\noindent where the quantum analog of the `light-tight box' condition is imposed: namely, that for the totality of all currents $j_j$, there are no initial or final states with real photons--i.e., no genuine photon `external lines.' (In the Davies theory, this amounts to the requirement that the existence of a real photon requires both an emitter and an absorber.) When only a subset of currents is considered, this condition is what yields the $D_1$ component as referenced above, giving rise to real `internal' photons corresponding to Fock states (but which are still tied to emitters and absorbers).\footnote{This is proved as a theorem in \cite{AB}, p. 302. The difference between $D_F$ and $\bar{D}$ vanishes when all currents are summed over, under the condition that there are no unsourced (`free') photons, which is an intrinsic aspect of the direct-action theory. This is discussed in detail in \cite{Davies71}.} 

With the replacement (5) for $A(x)$ (and with $J$ denoting the action and $T$ the time-ordering operator), the S-matrix becomes:\footnote{This is the S-matrix in terms of the action $J$, making use of the property that $S=Z[0]$ where $Z[j]$ is the generating functional of the path integral formulation, i.e.: $Z[j] \propto \int D\phi e^{i(J[\phi] + \int d^4x \phi(x) j(x) ) } $.}

$$ S = Te^{iJ} = T \  exp\  i \sum_i \sum_j {\frac{1}{2}  \int \int j_{(i)\mu}(x) D_F(x-y) j_{(j)}^{\mu}(y) d^4 x d^4 y}  \eqno (6) $$ 
 
\noindent where all currents $i$ and $j$ are summed over (this includes self-action for a given current $i$, and the factor of $\frac{1}{2}$ enters to compensate for double-counting of interacting distinguishable currents $i$ and $j$). 

We can now make use of standard results from quantum electrodynamics (QED) regarding `emission' and `absorption' processes; all we need to do is to recall that any occurrence of the field $A^{\mu}(x)$ represents the combined effect at $x$ of all currents $j_j$.  
This includes the Coulomb field (zeroth component, $A^0$), which is not quantized in the Davies theory, but we are interested in radiative phenomena (emission and absorption), corresponding to quanta of the field. For the latter to occur, we require a response to a current $j_i$ from at least one other current $j_j$ such that cancellation/reinforcement of the appropriate fields is achieved, thus creating a `free field' corresponding to a photon Fock state $|k\rangle$ (it is actually a projection operator as shown below).  Without the appropriately phased absorber response, we have only non-quantized virtual photons, represented by the time-symmetric propagator (first term on the right hand side of (4)), as opposed to Fock states (second term on the right hands side of (4)). A useful way to conceptualize this distinction is that `virtual photons are force carriers, but only real photons are energy carriers.'

The latter process--creation of a Fock state of momentum $k$ by way of the response of currents $j_j$ to the time-symmetric field of the emitting current $j_i$--corresponds functionally to the `action of a creation operator on the vacuum' in the usual quantized theory, i.e.:

$$ |k\rangle = \hat{a_k}^{\dag} |0\rangle   \eqno (7)  $$

The above characterizes the `emission of the photon' from the excited atom (i.e. the energy source), while the dual expression characterizes the generation of the CW by a responding current $j_j$:

$$ \langle k| = \langle 0| \hat{a_k}    \eqno(8) $$

This response characterizes not only the energy $k^0$ of the state, but a particular spatial momentum $\vec{k}$ corresponding to the relationship between the currents $j_i$ and $j_j$. By virtue of (7) and (8), and the fact that the field must be real-valued, we see that the real photon is most appropriately represented by a projection operator, $  | k \rangle \langle k|$, reflecting the fact that absorber response is the key component of the non-unitary measurement transition. The representation of the occurrence of a real photon through a sum of projection operators (as opposed to a ket) is also implicit in Davies' discussion of factorization of the S-matrix for the case of a `real internal photon'  being emitted at $y$ and absorbed at $x$ (\cite{Davies72}, eq. (19)):

$$iD_+(x-y) = \langle 0|A_\nu(x) A_\mu(y)|0\rangle = \sum_{\vec{k}} \langle 0|A_\nu(x) |\vec{k} \rangle 
\langle \vec{k} | A_\mu(y)| 0\rangle   \eqno(9)  $$

\noindent  The above is the first order contribution to the S-matrix, which describes the emission and absorption of one photon. Due to absorber responses, there is a matter of fact as to which current emits the photon and which currents are eligible to receive it, which is what converts the Feynman propagator $D_F$ into a single factorizable vacuum expectation value of a product of field operators, as above.  This is explicitly shown in \cite{Davies72}, pp. 1030-1.\footnote{For convenience, here we have transposed the indices so that $D_+(x-y)$ may be used, although of course this is equivalent to $-D_-(y-x)$ (cf. \cite{Davies71}, eqn. (23)). (We suppress polarization indices.) Davies does not attribute the distinguishability of the currents to absorber response, and instead assumes that the currents are distinguishable due to physical separation, which presents ambiguity issues, i.e.: how large a physical separation is required for distinguishability, and why? In contrast, a clear criterion for `distinguishability of currents' is available if this is understood as arising from absorber responses, which allow factorization as in (9), since they give rise to the `free field' $D_+$. Of course, absorber response does not always occur. When it does not occur, there is no transfer of energy, but only a virtual photon connection corresponding to the (nonquantized) time-symmetric propagator $\bar{D}$.}

 It is important to note that (9) yields a sum over squares of the vector potential component $\vec{A}_k$, corresponding to units of [energy] per [wave vector squared], and is real-valued, representing the transfer of real energy $\hbar k c$ (see also (12) below for explicit definition of $\vec{A}$). That is, the propagator--typically representing a complex amplitude (as in the virtual photon case in which unitarity is retained)--has been transformed into a sum over real quantities, each corresponding to the square of the field component $\vec{A}_k$ (which itself is only an amplitude). This reflects a non-unitary measurement transition from a ket to a sum over possible outcomes, each characterized by a real eigenvalue (in this case  $\frac{\hbar c}{k}$). It has gone unnoticed in the standard theory however, since that neglects any contribution of absorbers to the physical process of exchange of a photon, and it is only through the latter that we are allowed to describe the interaction by $D_+$ (which expresses the non-unitarity) rather than $D_F$ (which does not distinguish between real and virtual photons). Indeed, the non-unitarity of the S matrix for cases in which a subset of currents is considered, giving rise to the $D_+$ term, is explicit in the decomposition (4) which yields a complex action. Given (6), this implies a non-unitary S matrix (again, considering only a subset of all currents). When one takes into account that the $D_+$ term is only present due to absorber response, this shows explicitly that it is absorber response that breaks unitarity. 

\indent In general, many currents $j_j$ will respond, and this is reflected in the above sum over $\vec{k}$, (9).\footnote{Davies describes the sum over $\vec{k}$ as a formal quantity only, but in our interpretation, this sum corresponds to the mixed state representing a set of incipient transactions, as in eq. (2).}  But real absorption can only occur for one of the responding currents, in conformance with the quantization of radiation, so this is a non-unitary process; i.e., only one projection operator, representing the particular spatial momentum component actualized, `remains standing' as the real photon. As noted above, the advanced responses, which thus break the linearity of the Schr\"{o}dinger evolution, are not present in the usual theory. Taking them into account explains why a real photon capable of transferring energy is represented by a projection operator and has become localized to the particular absorber excited by it, rather than continuing to propagate along unitarily (as a ket $|k\rangle$ representing only an amplitude) and to enter into ever-increasing numbers of correlations (leading to a Schr\"{o}dinger's cat scenario). 

Thus, the process of absorption (arising from absorber responses) which precipitates the non-unitary `measurement' transition, or `Process 1' of von Neumann (\cite{VN}) in the transactional model \textit{is indeed clearly and unambiguously defined}: it occurs whenever there is a standard absorption-type transition in a bound state, such as an atom or molecule. It may be precisely quantified for the case of atomic electrons in terms of the usual QED field $A_\mu$, keeping in mind that $A_\mu$ is a stand-in for a given current's interaction with other currents, as in eqn. (5). Specifically, `absorption' occurs in TI whenever a bound state component absorbs a photon of momentum $k$ and polarization $\alpha$, thereby transitioning from a stationary state $A$ to a higher stationary state $B$, as described by the matrix element (cf. \cite{Sakurai} p. 37):

$$ \langle B; n_{k,\alpha} -1| H_{int} | A; n_{k,\alpha} \rangle  \eqno(10) $$

The interaction Hamilton $H_{int}$ in the above is expressed in terms of the usual QED field (to lowest order) as

$$ H_{int} = -\frac{e}{mc}\vec{A}(x,t) \cdot \vec{p}   \eqno(11)  $$

\noindent where use has been made of the transversality condition applying to radiated quanta, i.e., $\nabla \cdot \vec{A} = 0$, (cf. \cite{Sakurai}, p. 36), and $\vec{A}$ now arises from responses from charges to the basic direct-action connection, as in eqn. (5). By virtue of those responses, the field is real-valued, and in its quantized form is a Hermitian operator:

$$ \vec{A}(\vec{x},t) =   \frac{c}{\sqrt{V}} \sum_{\vec{k}, \alpha} \sqrt{\frac{\hbar}{2\omega_k}}  [\hat{a}_{\vec{k},\alpha}(0) \vec{\epsilon}^{(\alpha)} e^{i(\vec{k}\cdot x-\omega_k t)} + \hat{a}_{\vec{k},\alpha}^\dag(0) \vec{\epsilon}^{(\alpha)} e^{-i(\vec{k}\cdot x-\omega_k t)} ] \eqno(12) $$

\indent Thus, `absorption' in TI is just as in standard QED, except that the `quantum electromagnetic field' 
$\vec{A}$, whose existence makes possible the transitions between atomic and molecular states, is acknowledged as arising from the responses of other charges to the field from a given current. I.e., the `creation' and `annihilation' operators comprising the field $\vec{A}$ are stand-ins for responses from charged currents interacting with the emitting current in such a way as to give rise to the real field corresponding to a real photon. It is the set of responses that breaks the linearity of the Schr\"{o}dinger evolution. But of course, in computational situations we cannot possibly take into account all these potentially responding currents, so using the field $\vec{A}$ as a calculational device makes perfect sense. However, one can do that without taking $\vec{A}$ as the fundamental ontology, and this resolves consistency problems such as Haag's Theorem \cite{KastnerHaag}.

Although implied in eqn.(9), it is worth noting that the squaring procedure of the Born Rule is explicitly derived in the direct-action theory from the fact (discussed above) that radiative processes, i.e., processes involving real photon transfer, occur only when there is \textit{both} emission and absorption; neither is a unilateral process. Thus, an emission \textit{or} absorption by a given atom, as described by standard QED, is only half of the entire process. In the standard approach, when we calculate the amplitude for emission of a photon by an atom, we ignore absorption of that same photon by another atom; and vice versa. Indeed, it is assumed in the standard approach that no absorption need occur for emission, and vice versa (i.e. the existence of unsourced `free fields' is allowed, whereas it is prohibited in the direct-action theory). Then we need to square that amplitude (for either emission or absorption) to get the probability of the half of the process we are considering. If, instead, we calculate the amplitude for both processes together, we actually end up squaring the amplitude for either process, arriving at the Born Rule. Thus, the Born Rule naturally arises when emission and absorption are both required for the existence of a real photon.

Let us now do this calculation explicitly. Consider the amplitude for emission of a photon of frequency $\omega_k$ by atom E and absorption of that same photon by another atom A (which defines a particular wave vector $\vec{k}$). E is in an initial excited state of energy $\epsilon_1= \hbar \omega_1$ and A is in a lower energy state $\epsilon_0= \hbar \omega_0$. The difference in the atomic frequencies is $\omega_1 - \omega_0 = \Delta \omega$. From time-dependent perturbation theory we have the standard formula for either emission or absorption involving the relevant matrix elements; these are essentially (10) and (11), where the interaction Hamiltonian for emission involves only the creation operator ${\hat{a}^\dag}_k e^{-i(\vec{k}\cdot x-\omega_k  t}$ and that for absorption only the annihilation operator $\hat{a_k} e^{i(\vec{k}\cdot x-\omega_k t}$. We also have to integrate  with respect to time to get the transition amplitudes as a function of $t$.  The time dependence for the emission part is 

$$ \int_0^t d\tau e^{-i\Delta \omega\tau} e^{i\omega_k\tau } = \frac{e^{i (-\Delta \omega +\omega_k)t }-1}
{i (-\Delta \omega + \omega_k)}    \eqno(13) $$

\noindent and for the absorption part:

$$ \int_0^t d\tau e^{i\Delta \omega\tau} e^{-i\omega_k\tau } = \frac{e^{i (\Delta \omega -\omega_k)t }-1}
{i (\Delta \omega - \omega_k)}    \eqno(14) $$

Note that (14) is just the complex conjugate of (13), and the matrix elements are also complex conjugates of each other. Thus, the total time-dependent amplitude for the combined processes of emission \textit{and} absorption of a photon of frequency $\omega_k$ by atoms E and A respectively is: 

$$\langle \epsilon_1; 0| H_{int} | \epsilon_0; k \rangle  \langle \epsilon_0; k| {H_{int}}^\dag | \epsilon_1; 0 \rangle  {\left| \frac{e^{i (\Delta \omega -\omega_k)t }-1}
{i (\Delta \omega - \omega_k)}\right|}^2 =
{\left| \langle \epsilon_1; 0| H_{int} | \epsilon_0; k \rangle \right|}^2 {\left| \frac{e^{i (\Delta \omega -\omega_k)t }-1} {i (\Delta \omega - \omega_k)}\right|}^2
  \eqno(15) $$

\noindent which clearly turns out to be the \textit{probability} for emission \textit{or} absorption of a photon in mode $\vec{k}$. When one considers large values of $t$, the time-dependent factor becomes a delta function, enforcing energy conservation, and a decay (or excitation) rate for that mode is obtained for the emitting or absorbing atom respectively. Thus, we see that the direct-action theory, which requires that all radiative processes involve \textit{both} emission and absorption, naturally yields the squaring procedure of the Born Rule.

\indent One might still wonder: what is it that `causes' a charged current to respond to the basic time-symmetric field from another charge? While this is not a causal process--it is indeterministic--we can quantify it as follows. Feynman noted regarding QED that the coupling constant $e$ is the amplitude for a real photon to be emitted or absorbed.  Now, in the direct action theory, in order for a real photon to exist at all, both processes must occur; i.e., there must be \textit{both} a time-symmetric field emitted from some current $j_i$ \textit{and} a response from other current(s) $j_k$.  In this way, we obtain for the basic probability of creation and destruction of a real photon two factors of the coupling constant; i.e., the fine structure constant $\alpha \sim \frac {1}{137} \sim  0.007$. Thus, for any given interaction between elementary charges, the probability that real emission and absorption will occur is very small, and virtual (force-based) interactions are predominant. In addition, any radiative process will have to obey the relevant conservation laws for the systems in question, which further decreases the probability of its occurrence. However, an object comprising a very large number of ground state atoms exposed to a large number of excited atoms has a good chance of responding, since all that is required is that any one of its constituent atoms responds. Again, to trigger non-unitary collapse, the response must be a Fock state $\langle k|$, corresponding (in this case) to definite photon number 1, since that is the only way that the future-directed `free field' is created corresponding to `radiation reaction', i.e., the loss of energy from the radiating charge.

To avoid any confusion, an aside is probably in order regarding other states of the field. The above is in contrast to a coherent state $|\alpha \rangle$ in which photon number is indefinite. The coherent state $| \alpha \rangle $ is defined in terms of the Fock basis as $| \alpha \rangle \propto  e^{\alpha \hat{a}^\dag} |0\rangle $ where $\alpha$ is the amplitude of the coherent state. Coherent states are the closest approach of the quantum electromagnetic field to the classical field, preserving phase relations (in contrast to Fock states which lack phase information). They may be created and temporarily imprinted on systems such atoms, but that process involves keeping atoms in superpositions of stationary states, and there can be no localization of individual photons to specific atoms under such circumstances, since there is no definite status as to absorption. This is related to the conjugate relationship between phase and photon number. In terms of the direct-action theory, an atom used to `store' a coherent state does not provide the necessary response for localization of a photon, since the atom must be retained in a superposition of stationary states. Thus there can be no well-defined application of the creation/annihilation operators (as in eqs (7) and (8)) defining a transition from one stationary state to another. For were that the case, the atom would either definitely be excited to a higher stationary state, or for a null measurement (in which it responds but another current `wins' the actualized photon), remain in the lower stationary state. In either case, the superposition required for coherent state storage would be lost. The deep physical reason behind the uncertainty in photon number accompanying phase coherence (which requires precise time correlation) is that the time of emission of any individual photon (Fock state with precise energy) is undefined in view of the uncertainty principle. Thus, at any particular time, an amplitude may be defined (corresponding to phase), but not a number of photons. 

Returning now to emission and absorption of Fock states: we can identify $e$ as the amplitude for a confirming response from a current $j_k$ present as a possible absorber. This provides a quantitative measure of the likelihood of absorption. Of course, this process is still indeterministic in nature: there is no way to predict, for any individual current, whether it will respond or not. That is in keeping with the indeterministic quality of quantum theory and reflects a deeper, relativistic level of indeterminacy. Nevertheless, we can now quantify the circumstances of absorber response, which allows for identification of the typical scale at which the measurement transition takes place and allows for placement of the `Heisenberg Cut' at the appropriate microscopic (or possibly mesoscopic) level of absorption by individual atoms or molecules.

\section{Conclusion}

\indent  It is shown herein that emission and absorption processes are quantitatively well-defined in the transactional (direct-action) picture, and are essentially the same as in the standard theory of quantum electrodynamics, except for the replacement of the quantized field by the response of charged currents $j_j$ to an emitting current ${j_i}$. Such emissions and responses cannot be predicted--they are inherently indeterministic. But the physical circumstances of their occurrence can be defined and quantified by identifying the coupling constant between interacting fields ($e$ in the case of the electromagnetic interaction) as the amplitude for generation of an OW (Fock state $|k\rangle$) or CW (dual Fock state $\langle k|$ ), both being required for the existence of a `real photon,' which in the direct-action picture is described by a Fock state projection operator $|k\rangle \langle k|$ .  Virtual photons are identified as the basic time-symmetric connections or propagators between currents, which do not prompt responses, do not precipitate the non-unitary transition, and thus remain an aspect of unitary (force-based) interactions only. Thus, virtual photons (time-symmetric propagator) convey force only, while real photons (projection operators, quanta of a real-valued field) convey real energy and break linearity. The latter is just an expression of what Einstein noted long ago: real electromagnetic energy (the \textit{actualized} photon $|k\rangle \langle k|$ ) is emitted and absorbed as a particle (projection operator with definite spatial momentum $\vec {k}$) \cite{Einstein}. It has been shown herein that the product of the amplitudes of emission and absorption constitute the squaring process for obtaining the probability of either radiative process considered separately, thus demonstrating that the Born Rule arises naturally in the direct-action theory of fields, in which both processes must always occur together (i.e., there is never emission without absorption, and vice versa).

 \indent   Finally, any quantized field theory can be re-expressed as a direct action theory, as shown by Narlikar \cite{Narlikar}.  Therefore, any field for which the basic Davies model holds is a component of the transactional model, and transfers of real quanta of those fields can be understood as the result of actualized transactions. (However, there is an asymmetry between gauge boson fields and their fermionic sources, and in general such sources participate in transactions indirectly, by way of boson confirmations \cite{Kastner_Maudlin}).  While the direct-action theory has historically been regarded with distrust, it is perfectly self-consistent; and it should also be noted here that as recently as 2003, Wheeler himself was advocating reconsideration of the direct-action picture \cite{WesleyWheeler}.
 
\section{Acknowledgments.}
The authors appreciate helpful comments from an anonymous referee for improvement of the presentation.

\end{document}